\documentclass[prb,superscriptaddress,reprint,amsmath,amssymb,longbibliography]{revtex4-2}

\usepackage{bm}
\usepackage{epsfig}
\usepackage{graphicx}
\usepackage{color}
\usepackage[colorlinks]{hyperref}
\usepackage[english]{babel}
\usepackage{float}
\usepackage{empheq}
\usepackage[bbgreekl]{mathbbol}
\usepackage{units}
\usepackage[normalem]{ulem}
\usepackage{dcolumn}

\begin{document}

\author{A.~N. Kamenskii}
\affiliation{Experimentelle Physik 2, Technische Universit\"at Dortmund, 44221 Dortmund, Germany}

\author{G.~G. Kozlov}
\affiliation{Spin Optics Laboratory, St.\,Petersburg State University, Peterhof, 198504 St.\,Petersburg, Russia}

\author{E. I. Baibekov}
\affiliation{Kazan Federal University, 420008 Kazan, Russia}

\author{B. Z. Malkin}
\affiliation{Kazan Federal University, 420008 Kazan, Russia}

\author{M. Bayer}
\affiliation{Experimentelle Physik 2, Technische Universit\"at Dortmund, 44221 Dortmund, Germany}
\affiliation{Ioffe Institute, Russian Academy of Sciences, 194021 St.\,Petersburg, Russia}

\author{A. Greilich}
\affiliation{Experimentelle Physik 2, Technische Universit\"at Dortmund, 44221 Dortmund, Germany}

\author{V.~S. Zapasskii}
\affiliation{Spin Optics Laboratory, St.\,Petersburg State University, Peterhof, 198504 St.\,Petersburg, Russia}

\date{\today}

\title{Nonlinear Faraday effect and spin noise in rare-earth activated crystals}

\begin{abstract}
The spin-noise spectroscopy (SNS) method implies high efficiency of conversion of the spin-system magnetization to the Faraday rotation angle.  Generally, this efficiency cannot be estimated using the characteristics of the regular magneto-optical activity of a paramagnet. However, it may be drastically enhanced in systems with strong inhomogeneous broadening of the optical transitions. This enhancement leads to the {\it giant spin-noise gain effect} and previously allowed one to apply the SNS to rare-earth-activated crystals. We show that the {\it nonlinear resonant} Faraday effect can be used to measure the homogeneous width of the inhomogeneously broadened transition and, thus, to estimate the applicability of the SNS to this type of paramagnet. We present the theoretical description of the effect and perform measurements on intraconfigurational ($4f$-$4f$) transitions of the trivalent rare-earth ions of neodymium and ytterbium in fluorite-based crystals. The proposed experimental approach establishes new links between the effects of nonlinear optics and spin-noise characteristics of crystals with paramagnetic impurities and offers new ways of research in the physics of impurity crystals.
\end{abstract}

\maketitle

\section*{Introduction}

The method of detecting magnetic resonances in the noise of magneto-optical activity --  the so-called {\it spin-noise spectroscopy} (SNS) -- has been intensely developed during the last 15 years~\cite{Muller,vzap,rise,glaz}. Having been primarily demonstrated on an atomic system~\cite{AZ81}, this technique has gained extensive use since its successful application to semiconductors~\cite{Oestr}. In spite of its apparently lower sensitivity of detecting stochastic (rather than regular) signals, SNS proved to be highly efficient not only as a conceptually new method of radiospectroscopy, but also as a peculiar method of optical spectroscopy, with its abilities substantially exceeding those of conventional linear spectroscopy~\cite{glaz}.	

Until recently, all attempts to apply SNS with its unique potentialities to dielectrics with paramagnetic impurities had failed~\cite{Polarimetry} because of the smallness of the specific Faraday rotation (Faraday-rotation cross section~\cite{Giri}) in these systems. This result can be explained by the fact that the allowed transitions, in activated crystals, are usually strongly broadened, while spectrally narrow lines are mostly weak. Both these factors affect unfavorably the Faraday rotation (FR) per unit spin density and the FR noise power.

In Ref.~[\onlinecite{OSN}], it was shown, however, that the value of the FR noise (spin-noise) power detected under conditions of resonant probing is determined not only by the FR per unit spin density, but also strongly depends on the homogeneous linewidth, which is hidden inside the inhomogeneously broadened profile and cannot be revealed in linear optical spectroscopy. By neglecting this fact, one may significantly underestimate the noise signal for the following reason. When the inhomogeneously broadened system is probed by a monochromatic laser beam in the absorption region, the polarization noise signal is contributed only by the ions of the sample whose resonant frequencies coincide with that of the laser light to within the homogeneous width $\gamma$. The fraction of these ions with respect to their total number in the beam can be estimated as $\gamma/\Gamma$, (here, $\Gamma$ is the inhomogenous linewidth of the optical transition) and may be extremely small. Under these conditions, the relative spin fluctuations of these ions may strongly exceed those of all the ions in the laser beam. It was found that the gain factor that determines the ``enhancement'' of the FR noise power is equal to the ratio of the inhomogeneous linewidth to the homogeneous one and may cover several orders of magnitude.

In Ref.~[\onlinecite{Kamen}], this idea was successfully applied to crystals activated by rare-earth (RE) ions, with parity-forbidden intraconfigurational ($f$-$f$) transitions, whose inhomogeneous linewidths (usually lying in the range of several GHz) may exceed their homogeneous width by many orders of magnitude. This giant spin-noise gain effect has allowed us to observe, up to that point unreported, magnetic resonances in the FR noise spectrum of an impurity crystal.
At the same time, it was found, that among the chosen RE ions and chosen $f$-$f$ transitions, only some of them appeared to be amenable for detection of the ground-state magnetic resonance in the FR noise spectrum, while others, showing similar linear magneto-optics, did not reveal any noticeable FR noise~\cite{Kamen}. For further development of the SNS of impurity crystals, it is required to find out in more detail the laws of formation of the spin noise signal. As follows from our previous experimental results~\cite{OSN,Kamen}, reliable information about the applicability of SNS to an object cannot be obtained from its linear optical or magneto-optical properties. Therefore, we plan to establish a relation between the magnitude of the spin-noise power of a paramagnet and its {\it nonlinear} magneto-optical characteristics.

In this paper, we study {\it resonant} spectra of the nonlinear FR of RE-activated crystals in the range of the $f$-$f$ transitions and show that the diamagnetic contribution to the nonlinear FR spectrum (controlled only by the magnetic splitting of the energy levels) strongly depends on the probe beam intensity and dramatically differs for lines with different types of broadening. Thus, we have found that high-resolution spectroscopy of nonlinear FR can be used to measure, in a single-beam configuration, the homogeneous widths of inhomogeneously broadened transitions and thus to distinguish the $f$-$f$ transitions capable to reveal the giant spin-noise gain effect.

The paper is organized as follows: after the general motivation in Sec.~\ref{sec:I} we provide the theoretical background of the expected effect in Sec.~\ref{sec:II}; Sec.~\ref{sec:III} describes the experimental setup and studied samples, while in Sec.~\ref{sec:IV} we discuss the results of the measurements on different transition of RE ions; Sec.~\ref{sec:V} concludes the paper.

\section{General considerations and motivation}
\label{sec:I}

The FR spectra of paramagnetic ions in dielectric media are known to be controlled by two main contributions -- diamagnetic and paramagnetic~\cite{Buck,PP}. The first one is determined by the magnetic splitting of optical transitions, while the second is related to their intensity difference resulting from the Boltzmann distribution of populations over magnetic sublevels of the ground state. Correspondingly, the diamagnetic contribution is temperature-independent, while the paramagnetic one is proportional to the spin-system magnetization and, in the high-temperature limit, obeys Curie's law. Without entering into details of the FR spectra, we note that the diamagnetic contribution is described by the derivative of the line's dispersion curve and, therefore, increases with decreasing linewidth.

\begin{figure}
 	\begin{center}
 		\includegraphics[width=\linewidth]{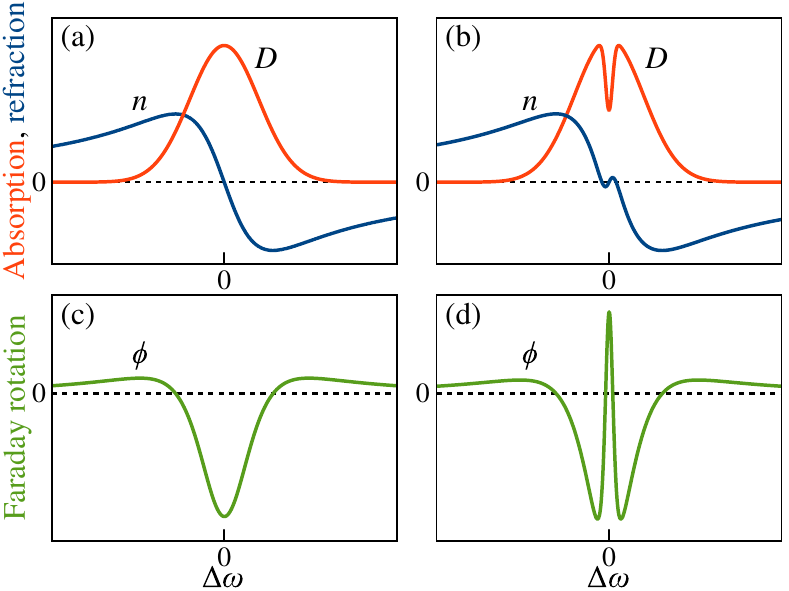}
 		\caption{(a) and (b) absorption ($D$) and refraction ($n$) spectra of an isolated inhomogeneously broadened optical transition; (b) for the transition with a spectral hole burnt by monochromatic laser light. (c) and (d) corresponding spectra of diamagnetic FR ($\phi$). The spectra of the transition with a spectral hole are supposed to be measured with an additional light source. $\Delta\omega$ is the relative detuning from the resonance.}
 		\label{Fig1}
 	\end{center}
\end{figure}

Along with the temperature and spectral properties of these two contributions, they can be distinguished by their inertial characteristics: the diamagnetic contribution responds to magnetic field variations practically instantaneously, while the paramagnetic contribution does it with some delay that is needed to establish thermal equilibrium of the populations over magnetic sublevels of the ground state. Thus, in the response to a magnetic field modulated at sufficiently high frequency (exceeding the ground-state spin-relaxation rate), the paramagnetic contribution can be strongly suppressed even at low temperatures~\cite{Mod}.

The behavior of the absorption and FR spectra of inhomogeneously broadened lines under conditions of strong resonant excitation has been previously studied for optical transitions in atomic (gaseous) systems, where the inhomogeneous broadening is associated with the Doppler effect~\cite{Bud1,Bud2}. In these systems, under usual experimental conditions, the magnetic splitting of the energy levels appears to be smaller than the thermal energy $k_\text{B}T$ ($k_\text{B}$ being the Boltzmann constant) by many orders of magnitude, and, as a result, the paramagnetic contribution to the Faraday effect may be neglected. In this case, the pure diamagnetic FR spectrum of an isolated optical transition looks as shown in Fig.~\ref{Fig1}(c).

Under conditions of strong resonant laser excitation of an inhomogeneously broadened transition, the absorption line profile (see Fig.~\ref{Fig1}(a)) becomes distorted by the hole-burning effect (Fig.~\ref{Fig1}(b)), and an additional spectral contribution (contribution of the 'hole') to the FR spectrum arises. As compared with the FR spectrum in the linear regime, this contribution has the opposite sign, is spectrally narrower, and, accordingly, is greater in magnitude (Fig.~\ref{Fig1}(d)). The ratio of the amplitudes of the two components ('broad' and 'narrow') to the FR spectrum, as can be expected, should correlate with the ratio of the inhomogeneous to homogeneous widths, which, as seen from the figure, can be measured in a single-beam (rather than pump-probe) configuration with the probe beam tuned to the line center.

As was shown in a number of previous publications~\cite{minerals,Mac} and noted in our recent work~\cite{Kamen}, this ratio, in the $f$-$f$ transitions of RE ions in crystals, may reach $6-8$ orders of magnitude, and, correspondingly, the nonlinear Faraday effect in these systems may exceed its linear counterpart also by many orders of magnitude. It is important to note that, in the single-beam measurements of the nonlinear Faraday effect, the wavelength of the laser beam always coincides with the hole center, where the nonlinear FR angle is the greatest. Thus, we can conclude that, with increasing intensity of the probe laser light, the resonant FR in the center of the $f$-$f$ transition will change its sign and increase in magnitude by many orders of magnitude.

\section{Theoretical background}
\label{sec:II}

This section presents a semi-quantitative consideration of the nonlinear Faraday effect observed for resonant probing of an inhomogeneously broadened spectral line. Similar phenomena in gas systems, with inhomogeneous broadening caused by the Doppler effect, were described in Refs.~[\onlinecite{Bud1,Bud2}]. Here, we consider optical transitions of paramagnetic ions in a crystal lattice, with a static inhomogeneous broadening associated with spatial fluctuations of the local crystal fields. The results of this treatment are further applied to RE-activated crystals. For consistency of their relation, we first consider the general characteristics of the FR spectra in the linear regime and, then, specific features of the nonlinear FR from inhomogeneously broadened transitions of ensembles of paramagnetic atoms (ions).

\subsection{Homogeneous broadening, linear FR}

\begin{figure}
 	\begin{center}
 		\includegraphics[width=0.5\linewidth]{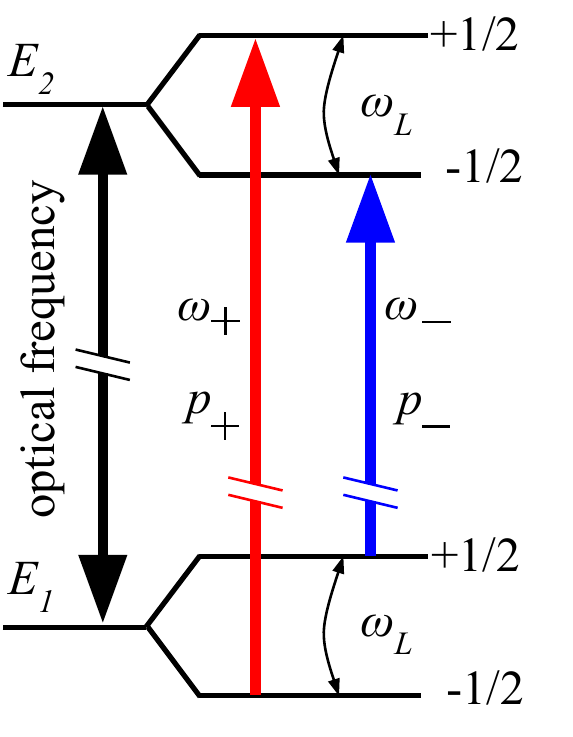}
 		\caption{The simplest energy-level diagram of a Kramers impurity ion in magnetic field. $E_1$ and $E_2$ are the ground and excited states, and $\pm 1/2$ indicate the spin levels split by the longitudinal magnetic field.  Under real experimental conditions, the magnetic splittings of the ground and excited states are much smaller than the total width of the transition, and the line profile, under strong resonant pumping, looks like curve $D$ in Fig.~\ref{Fig1}(a) or \ref{Fig1}(b) depending on whether the inhomogeneous width predominates homogeneous or not.}
 		\label{Fig2}
 	\end{center}
 \end{figure}

As is known, the FR angle $\phi$ is expressed in terms of the refractive indices $n_{\pm}(\omega)$ of the medium for the circularly polarized waves $\sigma_\pm$ as:
\begin{equation}
 \phi={\omega l\over c}\bigg [n_+(\omega)-n_-(\omega)\bigg ],
 \label{1}
\end{equation}
where $ \omega $ is the optical frequency of the probe beam, $ l $ is the sample length, and $ c $ is the speed of light. The contribution of the impurity ions to the refractive index $ n_+ (n_-) $ is determined by the transitions between the energy states, with the difference $ \Delta M $ of the projections of angular momentum on
the light propagation direction equal to $ +1 $ ($ -1 $) (Fig.~\ref{Fig2}).
In zero magnetic field, $B=0$, the values of the relative population difference $ p_\pm $ (the difference between the diagonal elements of the density matrix of the impurity system), as well as the frequencies $ \omega_\pm $
for the transitions with $ \Delta M = +1 $ and $ \Delta M = -1 $ are the same
(for zero magnetic field, we denote $ p_ \pm \equiv p $ and $ \omega_ \pm \equiv \omega_0 $), and the FR vanishes ($ \phi = 0 $).
When the magnetic field $ B $ is turned on, the energy levels of the impurity centers undergo the Zeeman splitting, which for the states with $ |\Delta M | = 1$  is equal to the Larmor frequency
$ \omega_L \equiv g \mu_\text{B} B / \hbar $ (here, $ g $ is the factor describing the magnetic splitting of a given ion, $ \mu_\text{B}$ is the Bohr magneton, and $ \hbar $ is the reduced Planck constant). Therefore, the frequencies and populations for the
transitions with $ \Delta M = + 1 $ and $ \Delta M = -1 $ become different and can be presented in the form:
\begin{equation}
 \omega_\pm=\omega_0\pm\omega_L,\hskip5mm p_\pm=p\pm \hbar\omega_L/2k_\text{B}T,
 \label{2}
 \end{equation}
where $ k_\text{B} $ is the Boltzmann constant, and $ T $ is the temperature. (Eq.~(\ref{2}) is valid for $ \hbar \omega_L / 2k_\text{B}T <1 $).
In the case of absence of the inhomogeneous broadening, when the frequency $\omega_0$ is the same for all impurity centers, we can represent the above refractive indices $n_\pm$ in the form:
 \begin{multline}
 n_\pm(\omega)= {2\pi \mathrm{d}^2 \over \hbar}\hskip1mm N  p_\pm\hskip1mm  f(\omega_\pm-\omega)=\\
 ={2\pi \mathrm{d}^2 \over \hbar}\hskip1mm N  [p\pm \hbar\omega_L/2k_\text{B}T]  f(\omega_0\pm \omega_L-\omega).
 \label{3}
 \end{multline}
 Here, $ \mathrm{d} $ is the dipole moment of the optical transitions $\sigma_\pm$,
 $ N $ is the concentration of the impurity centers,
 the function $ f $ describes the frequency dependence of the refractive indices $ n_ \pm (\omega) $, and for the
 homogeneously broadened optical transition has the form:
 \begin{equation}
 f(\nu) = {\nu \over\nu^2+\gamma^2},
 \end{equation}
 where $ \gamma $ is the width of the homogeneously broadened transition.
 In Eq.~(\ref{3}), we do not take into account the background part of the refractive index not associated with the considered transition of the impurity ion.

At low magnetic fields, where $ \omega_L <\gamma, k_\text{B}T $, the dependence of the FR angle on the magnetic field
(or on the Larmor frequency $ \omega_L $) is linear and can be found using Eq.~(\ref {1}):
 \begin{equation}
 {\phi\over \omega_L}={\omega l\over c}{d\over d\omega_L} \bigg [n_+(\omega)-n_-(\omega)\bigg ]\bigg |_{\omega_L=0}.
 \end{equation}
 Substituting here $ n_ \pm (\omega) $ from Eq.~(\ref {3}), we have:
\begin{align}
\phi & =\phi_C-\phi_A,\notag &\\
\phi_C & \equiv {V\hbar\over 2k_\text{B}T}f(\omega_0-\omega),
\hskip5mm \phi_A\equiv Vp{\partial \over \partial\omega}f(\omega_0-\omega),\notag &\\
V & \equiv {4\pi \mathrm{d}^2 N \over \hbar}\hskip1mm {\omega_L\omega l\over c }.&
\label{6}
\end{align}
The contribution $ \phi_C $,  usually referred to as {\it paramagnetic} or {\it C-term}~\cite{Buck} is associated with variations of the impurity energy-level populations in the magnetic field.
As seen from Eq.~(\ref{6}), the $ C $-term decreases with increasing temperature.
Dynamics of this contribution under variations of the magnetic field is controlled by the population relaxation time $ T_1 $,
so that, upon modulation of the magnetic field at frequencies exceeding $ 1 / T_1 $, the contribution $ \phi_C $ is being suppressed.
Thus, at sufficiently high temperatures ($ k_\text{B}T \gg \hbar \omega_L $) and under conditions of sufficiently high frequency of the magnetic field modulation, the contribution of the $ C $-term to the detected FR signal can be neglected.

The contribution $ \phi_A $, usually referred to as {\it diamagnetic} or {\it A-term}~\cite{Buck}, is related to the shift of the impurity energy levels in the applied magnetic field. This contribution, which is temperature-independent and practically inertialess, will be most important for interpretation of our experimental data.
Below, we consider the behavior of this contribution for inhomogeneously broadened transitions at high intensities of the probe beam, where the effects of optical nonlinearity become essential.

\subsection{Inhomogeneous broadening, nonlinear FR}

Let the line of the inhomogeneously broadened transition be centered at the frequency $ \bar \omega $ and described by the shape function $ {\cal P} (\nu)> 0$ $(\int {\cal P} (\nu) d \nu = 1 $), so that $ N {\cal P} (\omega_0- \bar \omega) d \omega_0 $ is the concentration of the impurity centers with the transition frequency lying within the range  $ [\omega_0, \omega_0 + d \omega_0] $. With the inhomogeneous broadening introduced in this way, the diamagnetic contribution to the FR is given by the relation:
\begin{equation}
\phi_A(\omega)=V\int d\omega_0 {\cal P}(\omega_0-\bar\omega)p(\omega_0){\partial \over \partial \omega}f(\omega_0-\omega).
\label{7}
\end{equation}
Equations~(\ref{3}) and (\ref{7}) show that the contribution $\phi_A(\omega)$ is proportional to the derivative of the refractive index $n_\pm$ (at ${\omega_L=0}$).

At low light intensity, when the impurity system is practically not perturbed by the light, the population difference $ p (\omega_0) $ for all spectral fractions remains the same ($ p (\omega_0) = p_ {eq} $), as well as the total population of the lowest states of the impurity centers. With increasing intensity, the population difference for the resonant and near-resonant spectral fractions may decrease, giving rise to the so-called hole-burning effect and to the nonlinear Faraday effect.

Let us estimate, first, the value of the FR in the linear regime. We assume here that the function $ {\cal P} (\nu) $ is essentially nonzero
within the range $ \nu \in [-\Gamma, \Gamma] $, where $ {\cal P} (\nu) \sim 1 / [2\Gamma] \equiv \bar {\cal P} $.
In this case, the quantity $ \Gamma $ characterizes the inhomogeneous width, which is assumed to be much greater than $\gamma $. Denoting the FR at
the center of the inhomogeneously broadened line ($\bar \omega $), in the absence of saturation, by $\phi_{A0}$ and using the accepted simplifications, we obtain for this quantity (setting $ \omega = \bar \omega $ in Eq.~(\ref{7})) the following estimate:
 \begin{equation}
 \phi_{A0}\approx -Vp_{eq}\bar{\cal P}\int_{\bar\omega -\Gamma}^{\bar\omega +\Gamma}
 {\partial \over \partial \omega_0}f(\omega_0-\bar\omega)\hskip1mm  d\omega_0=
 -{Vp_{eq}\bar{\cal P}\over 2\Gamma }.
 \label{8}
 \end{equation}

Let us make now a similar estimate for the FR $ \phi_ {A1} $ under the condition of optical saturation. As noted above, in this case, the function $ p (\omega_0) $ appears to be distorted by the 'hole' at the frequency $ \omega_0 \approx \bar \omega $.
We will characterize this hole by a bell-shaped function $ {\cal L} (\nu) $,
which reaches its maximum at $ \nu = 0 $, has a
width of $ \tilde \gamma> \gamma $, and is, in magnitude, restricted to $ 0 <{\cal L} (\nu) <1 $. Using this function, we can represent the population distribution $ p (\omega_0) $ in the form:
 \begin{equation}
 p(\omega_0)=p_{eq}[1-{\cal L}(\omega_0-\bar\omega)].
 \label{9}
 \end{equation}
As follows both from intuitive and from more rigorous theoretical considerations, the function $ {\cal L} (\nu) $, describing the hole, should meet certain requirements.
The depth of the hole $ {\cal L} (0) $ and its width $ \tilde \gamma $ should depend on the probe beam intensity $ I $, so that
$ \lim_ {I \rightarrow 0} {\cal L} (0) = 0 $ and $ \lim_ {I \rightarrow 0} \tilde \gamma = \gamma $.
With increasing intensity, the amplitude $ {\cal L} (0) $ should increase and tend to unity as $ I \rightarrow \infty $. Regarding the hole width $ \tilde \gamma $, at low and moderate intensities, when $ {\cal L} (0) \ll 1 $, it should be $ \sim \gamma $, and start to increase with further growth of the intensity.

Substituting Eq.~(\ref {9}) into Eq.~(\ref {7}) and taking into account that $ \Gamma \gg \tilde \gamma $ (i.e., the width of the burnt hole is always smaller than the inhomogeneous width), we can write for the FR angle $ \phi_ {A1} $ measured under these conditions the following relations:
\begin{align}
\phi_{A1} &= \phi_{A0}+ \notag \\
& + Vp_{eq}\int d\omega_0 {\cal P}(\omega_0-\bar\omega){\cal L}(\omega_0-\bar\omega)
{\partial \over \partial \omega_0} f(\omega_0-\bar\omega) \approx \notag \\
 & \approx \phi_{A0}+Vp_{eq} {\cal P}(0)\int d\omega_0 {\cal L}(\omega_0-\bar\omega)
{\partial \over \partial \omega_0} f(\omega_0-\bar\omega).
\label{10}
\end{align}

Now, taking into account that the function $ {\cal L} (\omega_0- \bar \omega) $ is essentially different from zero at
$ \omega_0 \in [\bar \omega- \tilde \gamma, \bar \omega + \tilde \gamma] $ (where it is $ \sim {\cal L} (0) $),
and also the fact that $ {\cal P} (0) \sim \bar {\cal P} $, we obtain, for $ \phi_ {A1} $, the following estimate:
 \begin{align}
 \phi_{A1} & \approx \phi_{A0} +Vp_{eq} \bar{\cal P}{\cal L}(0)
 \int_{\bar\omega-\tilde\gamma}^{\bar\omega+\tilde\gamma}d\omega_0
 {\partial \over \partial \omega_0} f(\omega_0-\bar\omega)\approx \notag \\
 & \approx Vp_{eq} \bar{\cal P}\bigg [{{\cal L}(0)\over 2\tilde\gamma }-{1\over 2\Gamma}\bigg ].
  \label{11}
\end{align}
In accordance with the properties of the quantities $ {\cal L} (0) $ and $ \tilde \gamma $ described above, with increasing intensity of the probe, the first fraction in the square brackets first increases (since $ {\cal L} (0) $ grows), reaches its maximum at a certain intensity $ I = I_c $ and then slowly decreases (when $ {\cal L} (0) \approx 1 $ and $ \tilde \gamma $ continues to grow). From Eq.~(\ref {11}) one can see that, at sufficiently high intensities, the first term becomes greater than the second, and $ \phi_ {A1} $ changes its sign.

The relationship (\ref {11}) allows us to propose a way to estimate the ratio $\Gamma/ \gamma$ (inhomogeneous width to homogeneous width), which is, as noted in Ref.~[\onlinecite{Kamen}], an important parameter for assessing the applicability of SNS to a particular impurity system. Indeed, when the probe beam intensity is equal to $ I_c $, the value of $ {\cal L} (0) $ is already close to unity, while the hole width $ \tilde \gamma $ is yet close to the homogeneous width $\gamma$. Now, as can be seen from Eqs.~(\ref {11}) and (\ref {8}), the easily measurable ratio $ R \equiv \phi_ {A1} / \phi_ {A0}$ can be estimated as:
 \begin{equation}
 R|_{I=I_c}= {\phi_{A1}|_{I=I_c}\over \phi _{A0}}\sim{\Gamma\over \gamma}.
 \label{12}
 \end{equation}

Thus, we see that the large value of the ratio of the inhomogeneous width of a transition to its homogeneous width provides not only the ``giant SN gain effect''~\cite{Kamen}, but may also give rise to a ``giant nonlinear FR''.

\begin{figure}
 	\begin{center}
 		\includegraphics[width=\linewidth]{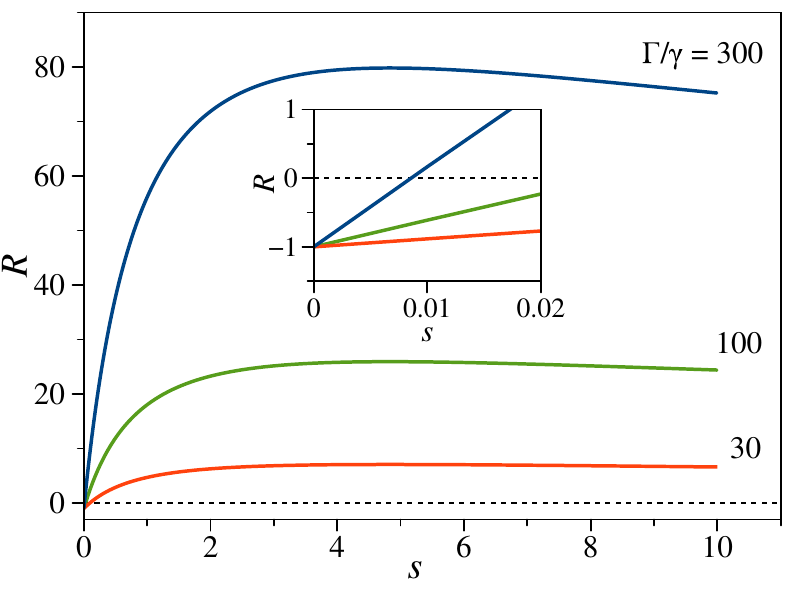}
 		\caption{Dependence of the resonant nonlinear FR (in units of linear FR) on the probe beam intensity (in units of the saturation factor $s$) for three values of the ratio $\Gamma/\gamma$. The inset in the center shows inversion of the FR sign at low light intensities.}
 		\label{Fig3}
 	\end{center}
 \end{figure}

A more rigorous quantitative consideration, which we do not present here, confirms the above reasoning and leads to the following formula for the ratio $ R $ introduced above:
\begin{align}
R &=  {\Gamma^2+\gamma^2\over \Gamma\gamma }{\pi s\over 2 \sqrt{s+1}\bigg (\sqrt{s+1}+1\bigg )^2}-1, \notag\\
s &\equiv {\Omega_\text{R}^2\over \gamma\gamma^{exc}}.
\label{13}
\end{align}
Here, $ \Omega_\text{R} \sim \sqrt I $ and $ \gamma^{exc}$ are the Rabi frequency and the excited-state decay rate, respectively, of the impurity-ion transition. An example of the dependence $R(s)$ is presented in Fig.~\ref{Fig3}. Qualitatively, the behavior of the nonlinear FR described by this formula is well correlated with our expectations: with increasing intensity of the probe beam, the FR, first, decreases, then changes its sign, before it strongly increases in magnitude.

The value $ s \sim I $ is usually referred to as the {\it saturation factor}.  The second fraction in Eq.~(\ref {13}) reaches its maximum (0.27) at $ s = s_c=4.81 $. The probe beam intensity (the light power density) corresponding to this value of the saturation factor was denoted above by $ I_c $.

Now, one can easily see that, at $ \Gamma \gg \gamma $, Eq.~(\ref {13}) yields a result similar to (but more accurate than) the estimate (\ref {12}):
\begin{equation}
R|_{I=I_c}=0.27 {\Gamma\over \gamma}.
\label{027}
\end{equation}
This equation can be used to estimate the homogeneous width of the optical transition from the "FR gain factor" and to predict the efficiency of application of the SNS technique to this particular system.

\section{Methods and Samples}
\label{sec:III}

To perform the experiments suggested above with a solid-state paramagnet, it is not enough to measure the FR at high intensity of the probe beam. One also has to get rid of the paramagnetic contribution, which usually predominates at low temperatures and may substantially distort the results of the measurements. In addition, the spectral width of the probe beam should be smaller than the homogeneous width of the transition under study.

\begin{figure}
\begin{center}
\includegraphics[width=\linewidth]{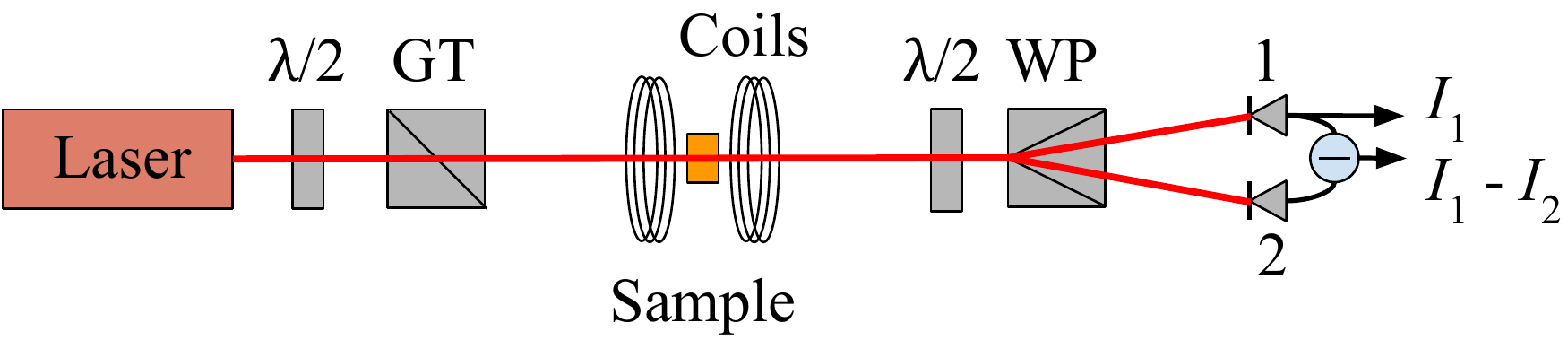}
\caption{A sketch of the experimental setup for measuring the FR angle. The laser beam power was controlled by the $\lambda/2$ plate and Glan-Taylor (GT) prism. Longitudinal magnetic field was created by a pair of coils. Polarimetric detector consisted of $\lambda/2$ plate, Wollaston prism (WP), and a pair of photo diodes. The FR angle was obtained by measuring the difference signal $I_1-I_2$ normalised by $I_1$.}
\label{Fig4}	
\end{center}
\end{figure}

The schematics of the experimental setup is presented in Fig.~\ref{Fig4}. As light source we used a tunable Ti:sapphire continuous wave ring laser (Coherent MBR110) with the linewidth of around 40\,kHz. The linearly polarized laser beam of 2\,mm diameter was focused at the sample using the 60\,mm focal-length lens. The rotation of the polarization plane of the transmitted beam was analyzed by the half-wave plate followed by the Wollaston prism and balanced photodiodes (Newport Nirvana 2007). Using additionally the single diode output, we were able to evaluate the absolute value of the rotation angle. The sample was mounted inside the continuous flow cryostat on a cold finger at the temperature of $\sim 6$\,K. To measure the diamagnetic contribution to the FR, we applied a longitudinal oscillating magnetic field using an electro magnet, with its frequency ($f_m=333$\,Hz) substantially exceeding the spin-lattice relaxation rate of the ground-state spin system. This frequency was then used as a reference for a lock-in amplifier to detect the signal.

To be able to compare the results of magneto-optical and spin-noise measurements, we used here the same crystals that were studied in our previous work on SNS~\cite{Kamen}. These include a SrF$_2$ crystal coactivated with Nd$^{3+}$ (0.5 mol \%) and Yb$^{3+}$ (0.15 mol \%), and a crystal of CaF$_2$:Nd$^{3+}$ (0.1 mol \%).

\section{Results of the measurements}
\label{sec:IV}

\subsection{Nonlinear FR spectra}

\begin{figure}
\begin{center}
\includegraphics[width=\linewidth]{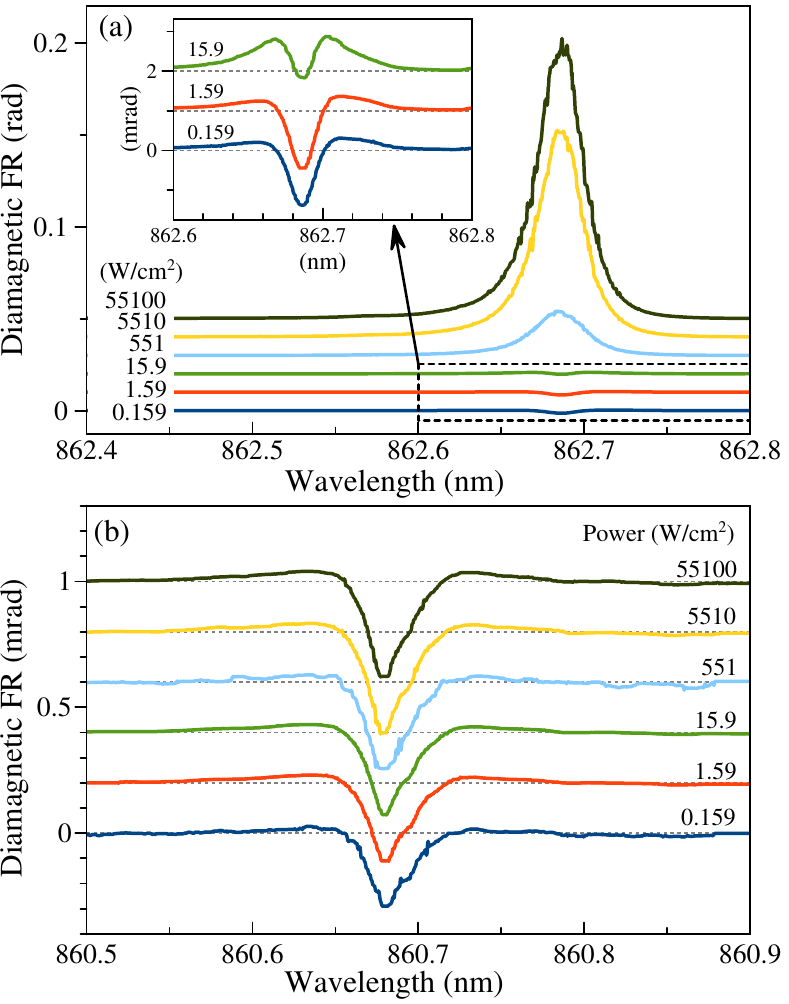}
\caption{Two patterns of behavior of the FR spectrum vs probe light intensity for inhomogeneously (a) and homogeneously (b) broadened transitions measured for the CaF$_2$-Nd (0.1\%) crystal. In the first case, panel (a), the FR, as a function of laser intensity, changes its sign at the resonance and grows in magnitude (see also the inset for lower powers), while in the second case, panel (b), it remains the same in a wide range of laser intensities. The spectra are shifted for clarity, whereby the zero level is marked by the dashed horizontal line. Note the scale difference between panels (a) in radian and (b) in milliradian.}
\label{Fig5}	
\end{center}
\end{figure}

The measurements of the FR spectra were performed on several lines of the $f$-$f$ transitions of Nd$^{3+}$ and Yb$^{3+}$ ions in the  CaF$_2$ and SrF$_2$ crystals. As expected, the behavior of the FR spectra strongly differed at different transitions. Some of them had a complicated structure which did not allow us to apply our simplified model. Still, there were several well isolated lines with a pronounced dependence of their FR spectra on the probe beam intensity. As an example, Fig.~\ref{Fig5}(a) shows the evolution of the FR spectrum of the line at 862.68\,nm of Nd$^{3+}$ in CaF$_2$ with increasing probe beam intensity. This transition of the tetragonal center is characterized by the longest lifetime of the excited state ($\sim 1.5$\,ms)~\cite{han1993}. Several FR spectra at the lowest light intensities (presumably in the linear regime) are shown in the inset. As we can see, the observed behavior perfectly agrees with the predictions of our model: the spectrum of the linear FR (at low light intensity) approximately corresponds to the derivative of the line's refractive index. Then, with increasing light intensity, the FR, in the line center, inverts its sign and strongly increases in magnitude. In this particular case, the greatest nonlinear FR exceeded the linear FR by nearly two orders of magnitude.

This is, however, not a common type of intensity-related behavior of the FR spectrum. Figure~\ref{Fig5}(b) shows another example of this behavior, when the FR spectrum virtually does not vary with the light intensity. In accordance with Ref.~[\onlinecite{han1993}], this transition belongs to the Nd$^{3+}$ M-center (pair center), characterized by fast cross-relaxation between the levels of the Nd ions comprising the center, which drastically shortens the excited-state lifetime (to 92\,$\mu$s). Under these conditions, the line appears to be broadened homogeneously, and no spectral hole is produced.

\subsection{Measuring the homogeneous widths}

\begin{figure}
\begin{center}
\includegraphics[width=\linewidth]{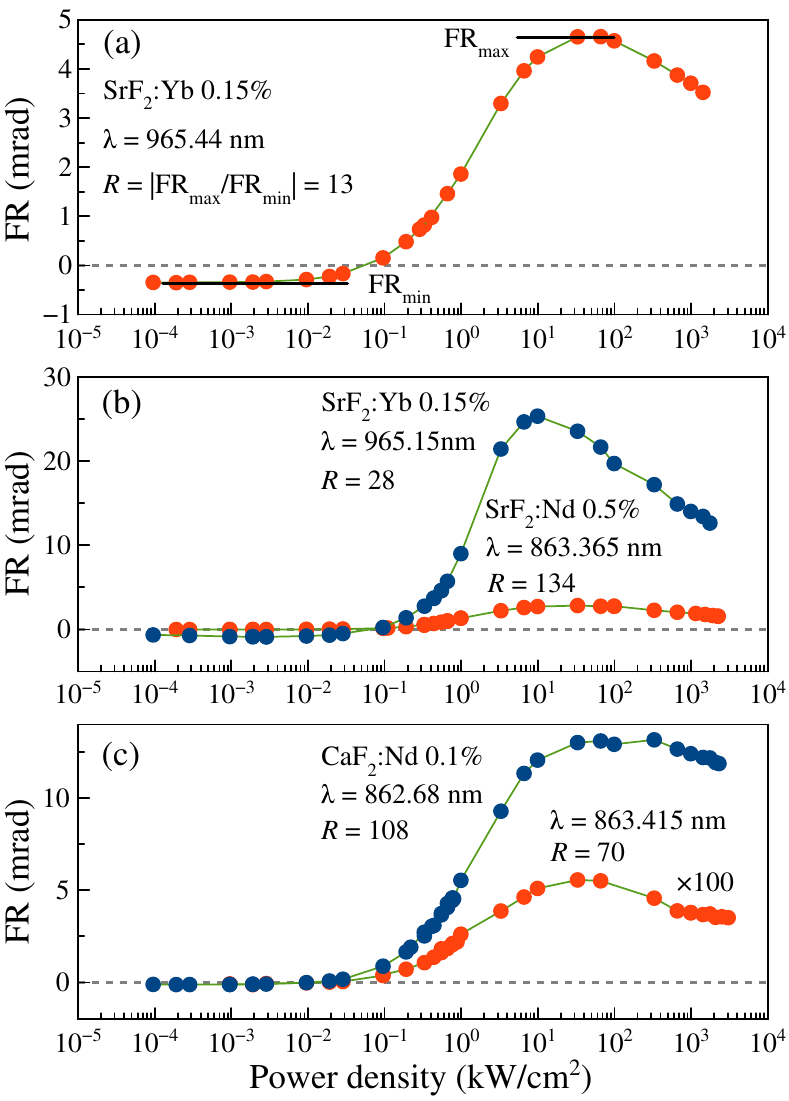}
\caption{Dependencies of the resonant FR for the studied transitions on the probe beam power density for SrF$_2$:Yb~(0.15\%) in the panel (a) and (b), for SrF$_2$:Nd~(0.5\%) in the panel (b), and for CaF$_2$:Nd~(0.1\%) in the panel (c). $\lambda$ is the probe wavelength at the transition. The FR enhancement factor $R=|\text{FR}_\text{max}/\text{FR}_\text{min}|$, see panel (a). $T = 6$\,K, maximal amplitude of the oscillating magnetic field $B_\sim = 0.3$\,mT, $f_m = 333$\,Hz. The green lines are guides for the eyes and are connecting the data points.}
\label{Fig6}		
\end{center}
\end{figure}

As follows from the above treatment, the homogeneous width of the transition can be found by measuring the factor of enhancement of the nonlinear FR ($R$). Figure~\ref{Fig6} shows the light-intensity dependencies of the resonant FR measured on several transitions of the RE ions in the studied crystals. The general pattern of these dependences, for all lines, correlates well with the results of our treatment presented in Fig.~\ref{Fig3}.

\begin{table}
\caption{Evaluation of the homogeneous width for some transitions of Nd$^{3+}$ and Yb$^{3+}$ in the studied crystals. $\lambda$ is the probe wavelength at the transition, $R$ is the FR enhancement factor, $\Gamma$ is the measured spectral width of the transition, and $\gamma$ is the homogeneous width of the transition, calculated using Eq.~(\ref{027}).}
\begin{tabular}{|l|l|l|l|c|}
\hline
Crystal & $\lambda$,\,nm & $R$ & $\Gamma$,\,GHz & $\gamma$,\,MHz \\
\hline\hline
SrF$_2$-Yb (0.15\%)& 965.44  & 13  & 2.68 & 44 \\
SrF$_2$-Yb (0.15\%)& 965.15  & 28  & 4.0  & 31 \\
SrF$_2$-Nd (0.5\%) & 863.365 & 134 & 26.8 & 54 \\
CaF$_2$-Nd (0.1\%) & 862.68 & 108 & 15.2 & 38 \\
CaF$_2$-Nd (0.1\%) & 863.415 & 70  & 12.8 & 50 \\
\hline
\end{tabular}
\label{table}
\end{table}

The results of the calculations of the homogeneous linewidths of the studied transitions obtained using our experimental data and Eq.~(\ref{027}) are presented in Table~\ref{table}. These results show that the quantity $\gamma$ is of the same order for all the transitions that revealed a pronounced nonlinear Faraday effect and lies in the range of a few tens of MHz. To make sure that the estimates of the homogeneous linewidths obtained from the measurements of the nonlinear FR are correct, we performed independent measurements of this quantity. The used method was based on the fact that the nonlinear FR measured in our experiments should depend linearly on the applied oscillating magnetic field only as long as the Zeeman splitting modulation remains smaller than the width of the hole burnt by the laser light in the inhomogeneously broadened line. Thus, the width of the hole
can be estimated by measuring the dependence of the FR signal on the applied field amplitude.

\begin{figure}
\begin{center}
\includegraphics[width=\linewidth]{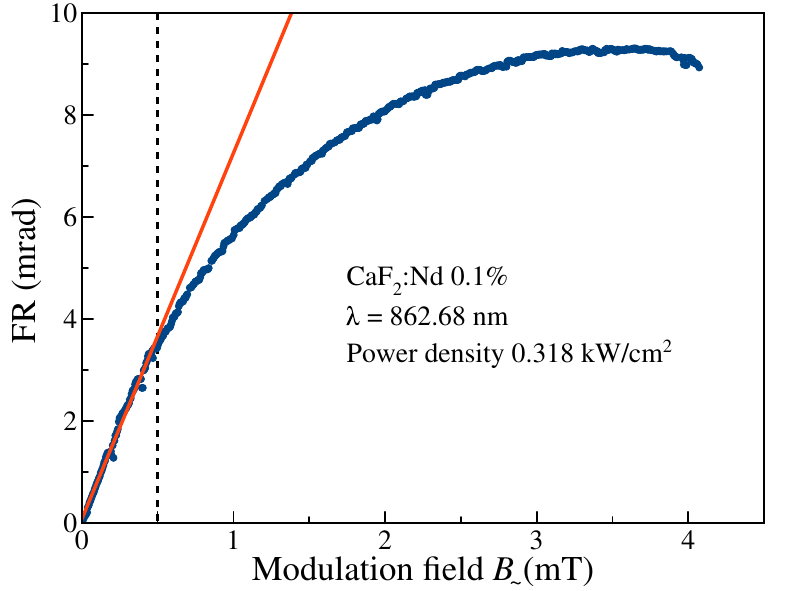}
\caption{Dependence of the FR signal on the amplitude of the applied oscillating magnetic field for the transition at 862.68\,nm of Nd$^{3+}$ in CaF$_2$. $T = 6$\,K, $f_m = 333$\,Hz. Red line is a linear fit. Vertical dashed line at 0.5\,mT shows the point where the data start to deviate from the linear dependence.}
\label{Fig7}
\end{center}
\end{figure}

Results of such measurements made on the 862.68\,nm line of Nd$^{3+}$ in CaF$_2$ are shown in Fig.~\ref{Fig7}. These measurements were performed at the light power density of around 0.3\,kW/cm$^2$, falling into the range of the initial linear dependence of the nonlinear FR on the probe beam intensity (Fig.~\ref{Fig6}), where the hole width is supposed to be not broadened by the light. It is also important that the amplitude of the oscillating field used in the measurements presented in Fig.~\ref{Fig6}, as one can see, belongs to the linear part of the plot in Fig.~\ref{Fig7}, which means that the approximation of smallness of the field amplitude, in these measurements, was satisfied.

As seen from Fig.~\ref{Fig7}, the presented dependence starts to deviate from a linear dependence at around 0.5\,mT, which, for the mean $g$ factor 3 yields the value $g \mu_\text{B} B/h \approx 21$\,MHz, correlating well with the measured homogeneous widths ($h$ is the Planck constant), see Table~\ref{table}.

\subsection{Discussion}

Now, we can check whether the results of the FR-based measurements of the homogeneous widths agree with the theoretical estimates of this value. To  evaluate the homogeneous width associated with the magnetic dipole-dipole interaction of the paramagnetic ions, we can use the following formula~\cite{Mim}:
\begin{equation}
\gamma = {\tilde k C \tilde g^2\mu_\text{B}^2\over \hbar},
\label{15g}
\end{equation}
where $\tilde g$ is the characteristic $g$ factor of the impurity ion, $\tilde k\sim 2.53$ is a numerical factor, governed by the relative values and signs of two pairs of $g$ factors of the ground and excited states of the ion, and $C$ is the impurity concentration. The estimate of $\gamma$ using Eq.~(\ref{15g}) for the tetragonal Nd$^{3+}$ center in the CaF$_2$ crystal with $C=2.46 \cdot 10^{19}$\,cm$^{-3}$ and $\tilde g=4$ yields $\gamma =2\pi \cdot 16$\,MHz, which, taking into account the uncertainties of our measurements and made approximations, agrees well with the experimental data of $\gamma$ for the transitions presented in Table~\ref{table}.

We did not find, in the literature, results of measuring the homogeneous linewidths of the $f$-$f$ transitions in these particular transitions, whereas it is known that, for other RE ions and for different $f$-$f$ transitions, these values may vary from tens of Hz to several GHz~\cite{Mac,book2005}. In particular, in Ref.~[\onlinecite{basiev}], the homogeneous width of the $f$-$f$ transitions of Nd$^{3+}$ in CaF$_2$ at 9\,K reached 350\,MHz. Studies of this kind are mainly aimed at searching for the smallest linewidths, most interesting from the viewpoint of applications~\cite{Appl0,Appl,Appl1}. For this study, we did not specially choose samples with an extraordinary small linewidth to observe a strong nonlinear FR, and we believe that, using other RE-doped crystals and other transitions, the effect of the enhanced nonlinear FR, being several orders of magnitude stronger, will become observable.

\section{Conclusions}
\label{sec:V}

In this paper, we have demonstrated specific features of the nonlinear resonant Faraday effect observed on the $f$-$f$ transitions of trivalent rare-earth ions in crystals. We show that, in transitions broadened essentially inhomogeneously and under conditions of sufficiently high intensity of the probe beam, the diamagnetic Faraday rotation may increase by several orders of magnitude as compared with its linear (unperturbed) value. In other words, we show that the giant spin-noise gain effect, not revealed in the regular {\it linear} Faraday rotation of the medium, can be revealed as an equally giant regular {\it nonlinear} Faraday effect under conditions of resonant probing. The results of our magneto-optical measurements confirm the correlation between the magnitude of the spin noise and the homogeneous width of the transition. The proposed magneto-optical method of testing optical transitions can be useful not only for studying the applicability of spin-noise spectroscopy to a particular system, but also as a single-beam method for measuring the homogeneous width of the optical transitions of paramagnetic impurities in crystals. Note also that the giant nonlinear Faraday rotation that may exceed the linear Macaluso-Corbino effect by many orders of magnitude, can find application in numerous magneto-optical devices for the laser wavelength stabilization and for magnetometric purposes~\cite{Bud2}.

\begin{acknowledgments}
We highly appreciate the financial support from the Deutsche Forschungsgemeinschaft in the frame of the International Collaborative Research Center TRR 160 (Project A5) and the Russian Foundation for Basic Research (Grant No. 19-52-12054). The authors from Russian side acknowledge the Saint Petersburg State University for the research Grant No. 73031758. E.I.B. acknowledges the support of the Foundation for the Advancement of Theoretical Physics and Mathematics "BASIS" and from the subsidy allocated to Kazan Federal University for the state assignment in the sphere of scientific activities No. FZSM-2020-0050.
\end{acknowledgments}


\begin{thebibliography}{24}%
\makeatletter
\providecommand \@ifxundefined [1]{%
 \@ifx{#1\undefined}
}%
\providecommand \@ifnum [1]{%
 \ifnum #1\expandafter \@firstoftwo
 \else \expandafter \@secondoftwo
 \fi
}%
\providecommand \@ifx [1]{%
 \ifx #1\expandafter \@firstoftwo
 \else \expandafter \@secondoftwo
 \fi
}%
\providecommand \natexlab [1]{#1}%
\providecommand \enquote  [1]{``#1''}%
\providecommand \bibnamefont  [1]{#1}%
\providecommand \bibfnamefont [1]{#1}%
\providecommand \citenamefont [1]{#1}%
\providecommand \href@noop [0]{\@secondoftwo}%
\providecommand \href [0]{\begingroup \@sanitize@url \@href}%
\providecommand \@href[1]{\@@startlink{#1}\@@href}%
\providecommand \@@href[1]{\endgroup#1\@@endlink}%
\providecommand \@sanitize@url [0]{\catcode `\\12\catcode `\$12\catcode
  `\&12\catcode `\#12\catcode `\^12\catcode `\_12\catcode `\%12\relax}%
\providecommand \@@startlink[1]{}%
\providecommand \@@endlink[0]{}%
\providecommand \url  [0]{\begingroup\@sanitize@url \@url }%
\providecommand \@url [1]{\endgroup\@href {#1}{\urlprefix }}%
\providecommand \urlprefix  [0]{URL }%
\providecommand \Eprint [0]{\href }%
\providecommand \doibase [0]{https://doi.org/}%
\providecommand \selectlanguage [0]{\@gobble}%
\providecommand \bibinfo  [0]{\@secondoftwo}%
\providecommand \bibfield  [0]{\@secondoftwo}%
\providecommand \translation [1]{[#1]}%
\providecommand \BibitemOpen [0]{}%
\providecommand \bibitemStop [0]{}%
\providecommand \bibitemNoStop [0]{.\EOS\space}%
\providecommand \EOS [0]{\spacefactor3000\relax}%
\providecommand \BibitemShut  [1]{\csname bibitem#1\endcsname}%
\let\auto@bib@innerbib\@empty
\bibitem [{\citenamefont {M{\"u}ller}\ \emph {et~al.}(2010)\citenamefont
  {M{\"u}ller}, \citenamefont {Oestreich}, \citenamefont {R{\"o}mer},\ and\
  \citenamefont {H{\"u}bner}}]{Muller}%
  \BibitemOpen
  \bibfield  {author} {\bibinfo {author} {\bibfnamefont {G.~M.}\ \bibnamefont
  {M{\"u}ller}}, \bibinfo {author} {\bibfnamefont {M.}~\bibnamefont
  {Oestreich}}, \bibinfo {author} {\bibfnamefont {M.}~\bibnamefont
  {R{\"o}mer}},\ and\ \bibinfo {author} {\bibfnamefont {J.}~\bibnamefont
  {H{\"u}bner}},\ }\bibfield  {title} {\bibinfo {title} {Semiconductor spin
  noise spectroscopy: Fundamentals, accomplishments, and challenges},\ }\href
  {https://doi.org/10.1016/j.physe.2010.08.010} {\bibfield  {journal} {\bibinfo
   {journal} {Physica E}\ }\textbf {\bibinfo {volume} {43}},\ \bibinfo {pages}
  {569} (\bibinfo {year} {2010})}\BibitemShut {NoStop}%
\bibitem [{\citenamefont {Zapasskii}(2013)}]{vzap}%
  \BibitemOpen
  \bibfield  {author} {\bibinfo {author} {\bibfnamefont {V.~S.}\ \bibnamefont
  {Zapasskii}},\ }\bibfield  {title} {\bibinfo {title} {Spin-noise
  spectroscopy: From proof of principle to applications},\ }\href
  {https://doi.org/10.1364/AOP.5.000131} {\bibfield  {journal} {\bibinfo
  {journal} {Adv. Opt. Photonics}\ }\textbf {\bibinfo {volume} {5}},\ \bibinfo
  {pages} {131} (\bibinfo {year} {2013})}\BibitemShut {NoStop}%
\bibitem [{\citenamefont {Hübner}\ \emph {et~al.}(2014)\citenamefont
  {Hübner}, \citenamefont {Berski}, \citenamefont {Dahbashi},\ and\
  \citenamefont {Oestreich}}]{rise}%
  \BibitemOpen
  \bibfield  {author} {\bibinfo {author} {\bibfnamefont {J.}~\bibnamefont
  {Hübner}}, \bibinfo {author} {\bibfnamefont {F.}~\bibnamefont {Berski}},
  \bibinfo {author} {\bibfnamefont {R.}~\bibnamefont {Dahbashi}},\ and\
  \bibinfo {author} {\bibfnamefont {M.}~\bibnamefont {Oestreich}},\ }\bibfield
  {title} {\bibinfo {title} {The rise of spin noise spectroscopy in
  semiconductors: From acoustic to {GHz} frequencies},\ }\href
  {https://doi.org/https://doi.org/10.1002/pssb.201350291} {\bibfield
  {journal} {\bibinfo  {journal} {phys. status solidi b}\ }\textbf {\bibinfo
  {volume} {251}},\ \bibinfo {pages} {1824} (\bibinfo {year}
  {2014})}\BibitemShut {NoStop}%
\bibitem [{\citenamefont {Glazov}(2016)}]{glaz}%
  \BibitemOpen
  \bibfield  {author} {\bibinfo {author} {\bibfnamefont {M.~M.}\ \bibnamefont
  {Glazov}},\ }\bibfield  {title} {\bibinfo {title} {Spin fluctuations of
  nonequilibrium electrons and excitons in semiconductors},\ }\href
  {https://doi.org/10.1134/S1063776116030067} {\bibfield  {journal} {\bibinfo
  {journal} {J. Exp. Theor. Phys.}\ }\textbf {\bibinfo {volume} {122}},\
  \bibinfo {pages} {472} (\bibinfo {year} {2016})}\BibitemShut {NoStop}%
\bibitem [{\citenamefont {Aleksandrov}\ and\ \citenamefont
  {Zapasskii}(1981)}]{AZ81}%
  \BibitemOpen
  \bibfield  {author} {\bibinfo {author} {\bibfnamefont {E.~B.}\ \bibnamefont
  {Aleksandrov}}\ and\ \bibinfo {author} {\bibfnamefont {V.~S.}\ \bibnamefont
  {Zapasskii}},\ }\bibfield  {title} {\bibinfo {title} {Magnetic resonance in
  the faraday-rotation noise spectrum},\ }\href@noop {} {\bibfield  {journal}
  {\bibinfo  {journal} {Zh. Eksp. Teor. Fiz. [Sov. Phys. JETP 54, 64 (1981)]}\
  }\textbf {\bibinfo {volume} {81}},\ \bibinfo {pages} {132} (\bibinfo {year}
  {1981})}\BibitemShut {NoStop}%
\bibitem [{\citenamefont {Oestreich}\ \emph {et~al.}(2005)\citenamefont
  {Oestreich}, \citenamefont {R\"omer}, \citenamefont {Haug},\ and\
  \citenamefont {H\"agele}}]{Oestr}%
  \BibitemOpen
  \bibfield  {author} {\bibinfo {author} {\bibfnamefont {M.}~\bibnamefont
  {Oestreich}}, \bibinfo {author} {\bibfnamefont {M.}~\bibnamefont {R\"omer}},
  \bibinfo {author} {\bibfnamefont {R.~J.}\ \bibnamefont {Haug}},\ and\
  \bibinfo {author} {\bibfnamefont {D.}~\bibnamefont {H\"agele}},\ }\bibfield
  {title} {\bibinfo {title} {Spin noise spectroscopy in {GaAs}},\ }\href
  {https://doi.org/10.1103/PhysRevLett.95.216603} {\bibfield  {journal}
  {\bibinfo  {journal} {Phys. Rev. Lett.}\ }\textbf {\bibinfo {volume} {95}},\
  \bibinfo {pages} {216603} (\bibinfo {year} {2005})}\BibitemShut {NoStop}%
\bibitem [{\citenamefont {Zapasskii}(2019)}]{Polarimetry}%
  \BibitemOpen
  \bibfield  {author} {\bibinfo {author} {\bibfnamefont {V.~S.}\ \bibnamefont
  {Zapasskii}},\ }\bibfield  {title} {\bibinfo {title} {Polarimetry of
  {{Regular}} and {{Stochastic Signals}} in {{Magnetooptics}}},\ }\href
  {https://doi.org/10.1134/S106378341905038X} {\bibfield  {journal} {\bibinfo
  {journal} {Phys. Solid State}\ }\textbf {\bibinfo {volume} {61}},\ \bibinfo
  {pages} {847} (\bibinfo {year} {2019})}\BibitemShut {NoStop}%
\bibitem [{\citenamefont {Giri}\ \emph {et~al.}(2012)\citenamefont {Giri},
  \citenamefont {Cronenberger}, \citenamefont {Vladimirova}, \citenamefont
  {Scalbert}, \citenamefont {Kavokin}, \citenamefont {Glazov}, \citenamefont
  {Nawrocki}, \citenamefont {Lema{\^i}tre},\ and\ \citenamefont
  {Bloch}}]{Giri}%
  \BibitemOpen
  \bibfield  {author} {\bibinfo {author} {\bibfnamefont {R.}~\bibnamefont
  {Giri}}, \bibinfo {author} {\bibfnamefont {S.}~\bibnamefont {Cronenberger}},
  \bibinfo {author} {\bibfnamefont {M.}~\bibnamefont {Vladimirova}}, \bibinfo
  {author} {\bibfnamefont {D.}~\bibnamefont {Scalbert}}, \bibinfo {author}
  {\bibfnamefont {K.~V.}\ \bibnamefont {Kavokin}}, \bibinfo {author}
  {\bibfnamefont {M.~M.}\ \bibnamefont {Glazov}}, \bibinfo {author}
  {\bibfnamefont {M.}~\bibnamefont {Nawrocki}}, \bibinfo {author}
  {\bibfnamefont {A.}~\bibnamefont {Lema{\^i}tre}},\ and\ \bibinfo {author}
  {\bibfnamefont {J.}~\bibnamefont {Bloch}},\ }\bibfield  {title} {\bibinfo
  {title} {Giant photoinduced {{Faraday}} rotation due to the spin-polarized
  electron gas in an $n$-{{GaAs}} microcavity},\ }\href
  {https://doi.org/10.1103/PhysRevB.85.195313} {\bibfield  {journal} {\bibinfo
  {journal} {Phys. Rev. B}\ }\textbf {\bibinfo {volume} {85}},\ \bibinfo
  {pages} {195313} (\bibinfo {year} {2012})}\BibitemShut {NoStop}%
\bibitem [{\citenamefont {Zapasskii}\ \emph {et~al.}(2013)\citenamefont
  {Zapasskii}, \citenamefont {Greilich}, \citenamefont {Crooker}, \citenamefont
  {Li}, \citenamefont {Kozlov}, \citenamefont {Yakovlev}, \citenamefont
  {Reuter}, \citenamefont {Wieck},\ and\ \citenamefont {Bayer}}]{OSN}%
  \BibitemOpen
  \bibfield  {author} {\bibinfo {author} {\bibfnamefont {V.~S.}\ \bibnamefont
  {Zapasskii}}, \bibinfo {author} {\bibfnamefont {A.}~\bibnamefont {Greilich}},
  \bibinfo {author} {\bibfnamefont {S.~A.}\ \bibnamefont {Crooker}}, \bibinfo
  {author} {\bibfnamefont {Y.}~\bibnamefont {Li}}, \bibinfo {author}
  {\bibfnamefont {G.~G.}\ \bibnamefont {Kozlov}}, \bibinfo {author}
  {\bibfnamefont {D.~R.}\ \bibnamefont {Yakovlev}}, \bibinfo {author}
  {\bibfnamefont {D.}~\bibnamefont {Reuter}}, \bibinfo {author} {\bibfnamefont
  {A.~D.}\ \bibnamefont {Wieck}},\ and\ \bibinfo {author} {\bibfnamefont
  {M.}~\bibnamefont {Bayer}},\ }\bibfield  {title} {\bibinfo {title} {Optical
  {{Spectroscopy}} of {{Spin Noise}}},\ }\href
  {https://doi.org/10.1103/PhysRevLett.110.176601} {\bibfield  {journal}
  {\bibinfo  {journal} {Phys. Rev. Lett.}\ }\textbf {\bibinfo {volume} {110}},\
  \bibinfo {pages} {176601} (\bibinfo {year} {2013})}\BibitemShut {NoStop}%
\bibitem [{\citenamefont {Kamenskii}\ \emph {et~al.}(2020)\citenamefont
  {Kamenskii}, \citenamefont {Greilich}, \citenamefont {Ryzhov}, \citenamefont
  {Kozlov}, \citenamefont {Bayer},\ and\ \citenamefont {Zapasskii}}]{Kamen}%
  \BibitemOpen
  \bibfield  {author} {\bibinfo {author} {\bibfnamefont {A.~N.}\ \bibnamefont
  {Kamenskii}}, \bibinfo {author} {\bibfnamefont {A.}~\bibnamefont {Greilich}},
  \bibinfo {author} {\bibfnamefont {I.~I.}\ \bibnamefont {Ryzhov}}, \bibinfo
  {author} {\bibfnamefont {G.~G.}\ \bibnamefont {Kozlov}}, \bibinfo {author}
  {\bibfnamefont {M.}~\bibnamefont {Bayer}},\ and\ \bibinfo {author}
  {\bibfnamefont {V.~S.}\ \bibnamefont {Zapasskii}},\ }\bibfield  {title}
  {\bibinfo {title} {Giant spin-noise gain enables magnetic resonance
  spectroscopy of impurity crystals},\ }\href
  {https://doi.org/10.1103/PhysRevResearch.2.023317} {\bibfield  {journal}
  {\bibinfo  {journal} {Phys. Rev. Research}\ }\textbf {\bibinfo {volume}
  {2}},\ \bibinfo {pages} {023317} (\bibinfo {year} {2020})}\BibitemShut
  {NoStop}%
\bibitem [{\citenamefont {Buckingham}\ and\ \citenamefont
  {Stephens}(1966)}]{Buck}%
  \BibitemOpen
  \bibfield  {author} {\bibinfo {author} {\bibfnamefont {A.~D.}\ \bibnamefont
  {Buckingham}}\ and\ \bibinfo {author} {\bibfnamefont {P.~J.}\ \bibnamefont
  {Stephens}},\ }\bibfield  {title} {\bibinfo {title} {Magnetic {{Optical
  Activity}}},\ }\href {https://doi.org/10.1146/annurev.pc.17.100166.002151}
  {\bibfield  {journal} {\bibinfo  {journal} {Annu. Rev. Phys. Chem.}\ }\textbf
  {\bibinfo {volume} {17}},\ \bibinfo {pages} {399} (\bibinfo {year}
  {1966})}\BibitemShut {NoStop}%
\bibitem [{\citenamefont {Zapasskii}\ and\ \citenamefont
  {Feofilov}(1975)}]{PP}%
  \BibitemOpen
  \bibfield  {author} {\bibinfo {author} {\bibfnamefont {V.~S.}\ \bibnamefont
  {Zapasskii}}\ and\ \bibinfo {author} {\bibfnamefont {P.~P.}\ \bibnamefont
  {Feofilov}},\ }\bibfield  {title} {\bibinfo {title} {Development of
  polarization magneto-optics of paramagnetic crystals},\ }\href
  {https://doi.org/10.1070/PU1975v018n05ABEH001959} {\bibfield  {journal}
  {\bibinfo  {journal} {Sov. Phys. Uspekhi}\ }\textbf {\bibinfo {volume}
  {18}},\ \bibinfo {pages} {323} (\bibinfo {year} {1975})}\BibitemShut
  {NoStop}%
\bibitem [{\citenamefont {Aleksandrov}\ and\ \citenamefont
  {Zapasskii}(1978)}]{Mod}%
  \BibitemOpen
  \bibfield  {author} {\bibinfo {author} {\bibfnamefont {E.}~\bibnamefont
  {Aleksandrov}}\ and\ \bibinfo {author} {\bibfnamefont {V.}~\bibnamefont
  {Zapasskii}},\ }\bibfield  {title} {\bibinfo {title} {Modulation
  magneto-optical spectroscopy of cross-relaxation resonances},\ }\href@noop {}
  {\bibfield  {journal} {\bibinfo  {journal} {Sov. Phys. Solid State}\ }\textbf
  {\bibinfo {volume} {20}},\ \bibinfo {pages} {679} (\bibinfo {year}
  {1978})}\BibitemShut {NoStop}%
\bibitem [{\citenamefont {Budker}\ \emph
  {et~al.}(2002{\natexlab{a}})\citenamefont {Budker}, \citenamefont {Kimball},
  \citenamefont {Rochester},\ and\ \citenamefont {Yashchuk}}]{Bud1}%
  \BibitemOpen
  \bibfield  {author} {\bibinfo {author} {\bibfnamefont {D.}~\bibnamefont
  {Budker}}, \bibinfo {author} {\bibfnamefont {D.~F.}\ \bibnamefont {Kimball}},
  \bibinfo {author} {\bibfnamefont {S.~M.}\ \bibnamefont {Rochester}},\ and\
  \bibinfo {author} {\bibfnamefont {V.~V.}\ \bibnamefont {Yashchuk}},\
  }\bibfield  {title} {\bibinfo {title} {Nonlinear electro- and magneto-optical
  effects related to {{Bennett}} structures},\ }\href
  {https://doi.org/10.1103/PhysRevA.65.033401} {\bibfield  {journal} {\bibinfo
  {journal} {Phys. Rev. A}\ }\textbf {\bibinfo {volume} {65}},\ \bibinfo
  {pages} {033401} (\bibinfo {year} {2002}{\natexlab{a}})}\BibitemShut
  {NoStop}%
\bibitem [{\citenamefont {Budker}\ \emph
  {et~al.}(2002{\natexlab{b}})\citenamefont {Budker}, \citenamefont {Gawlik},
  \citenamefont {Kimball}, \citenamefont {Rochester}, \citenamefont
  {Yashchuk},\ and\ \citenamefont {Weis}}]{Bud2}%
  \BibitemOpen
  \bibfield  {author} {\bibinfo {author} {\bibfnamefont {D.}~\bibnamefont
  {Budker}}, \bibinfo {author} {\bibfnamefont {W.}~\bibnamefont {Gawlik}},
  \bibinfo {author} {\bibfnamefont {D.~F.}\ \bibnamefont {Kimball}}, \bibinfo
  {author} {\bibfnamefont {S.~M.}\ \bibnamefont {Rochester}}, \bibinfo {author}
  {\bibfnamefont {V.~V.}\ \bibnamefont {Yashchuk}},\ and\ \bibinfo {author}
  {\bibfnamefont {A.}~\bibnamefont {Weis}},\ }\bibfield  {title} {\bibinfo
  {title} {Resonant nonlinear magneto-optical effects in atoms},\ }\href
  {https://doi.org/10.1103/RevModPhys.74.1153} {\bibfield  {journal} {\bibinfo
  {journal} {Rev. Mod. Phys.}\ }\textbf {\bibinfo {volume} {74}},\ \bibinfo
  {pages} {1153} (\bibinfo {year} {2002}{\natexlab{b}})}\BibitemShut {NoStop}%
\bibitem [{\citenamefont {Marfunin}(1979)}]{minerals}%
  \BibitemOpen
  \bibfield  {author} {\bibinfo {author} {\bibfnamefont {A.~S.}\ \bibnamefont
  {Marfunin}},\ }\href@noop {} {\emph {\bibinfo {title} {Spectroscopy,
  {{Luminescence}} and {{Radiation Centers}} in {{Minerals}}}}},\ \bibinfo
  {edition} {1st}\ ed.\ (\bibinfo  {publisher} {{Springer, Berlin,
  Heidelberg}},\ \bibinfo {year} {1979})\BibitemShut {NoStop}%
\bibitem [{\citenamefont {Macfarlane}(2002)}]{Mac}%
  \BibitemOpen
  \bibfield  {author} {\bibinfo {author} {\bibfnamefont {R.~M.}\ \bibnamefont
  {Macfarlane}},\ }\bibfield  {title} {\bibinfo {title} {High-resolution laser
  spectroscopy of rare-earth doped insulators: A personal perspective},\ }\href
  {https://doi.org/10.1016/S0022-2313(02)00450-7} {\bibfield  {journal}
  {\bibinfo  {journal} {J. Lumin.}\ }\textbf {\bibinfo {volume} {100}},\
  \bibinfo {pages} {1} (\bibinfo {year} {2002})}\BibitemShut {NoStop}%
\bibitem [{\citenamefont {Han}\ \emph {et~al.}(1993)\citenamefont {Han},
  \citenamefont {Jones},\ and\ \citenamefont {Syme}}]{han1993}%
  \BibitemOpen
  \bibfield  {author} {\bibinfo {author} {\bibfnamefont {T.~P.~J.}\
  \bibnamefont {Han}}, \bibinfo {author} {\bibfnamefont {G.~D.}\ \bibnamefont
  {Jones}},\ and\ \bibinfo {author} {\bibfnamefont {R.~W.~G.}\ \bibnamefont
  {Syme}},\ }\bibfield  {title} {\bibinfo {title} {Site-selective spectroscopy
  of {Nd}$^{3+}$ centers in {CaF}$_2$:{Nd}$^{3+}$ and {SrF}$_2$:{Nd}$^{3+}$},\
  }\href {https://doi.org/10.1103/PhysRevB.47.14706} {\bibfield  {journal}
  {\bibinfo  {journal} {Phys. Rev. B}\ }\textbf {\bibinfo {volume} {47}},\
  \bibinfo {pages} {14706} (\bibinfo {year} {1993})}\BibitemShut {NoStop}%
\bibitem [{\citenamefont {Mims}(1968)}]{Mim}%
  \BibitemOpen
  \bibfield  {author} {\bibinfo {author} {\bibfnamefont {W.~B.}\ \bibnamefont
  {Mims}},\ }\bibfield  {title} {\bibinfo {title} {Phase {{Memory}} in
  {{Electron Spin Echoes}}, {{Lattice Relaxation Effects}} in {CaWO}$_4$: {Er},
  {Ce}, {Mn}},\ }\href {https://doi.org/10.1103/PhysRev.168.370} {\bibfield
  {journal} {\bibinfo  {journal} {Phys. Rev.}\ }\textbf {\bibinfo {volume}
  {168}},\ \bibinfo {pages} {370} (\bibinfo {year} {1968})}\BibitemShut
  {NoStop}%
\bibitem [{\citenamefont {Liu}\ and\ \citenamefont
  {Jacquier}(2005)}]{book2005}%
  \BibitemOpen
  \bibinfo {editor} {\bibfnamefont {G.}~\bibnamefont {Liu}}\ and\ \bibinfo
  {editor} {\bibfnamefont {B.}~\bibnamefont {Jacquier}},\ eds.,\ \href@noop {}
  {\emph {\bibinfo {title} {Spectroscopic {{Properties}} of {{Rare Earths}} in
  {{Optical Materials}}}}},\ \bibinfo {edition} {1st}\ ed.,\ Springer
  {{Series}} in {{Materials Science}}\ (\bibinfo  {publisher} {{Springer,
  Berlin, Heidelberg}},\ \bibinfo {year} {2005})\BibitemShut {NoStop}%
\bibitem [{\citenamefont {Basiev}\ \emph {et~al.}(1998)\citenamefont {Basiev},
  \citenamefont {Karasik}, \citenamefont {Fedorov},\ and\ \citenamefont
  {Ver~Steeg}}]{basiev}%
  \BibitemOpen
  \bibfield  {author} {\bibinfo {author} {\bibfnamefont {T.~T.}\ \bibnamefont
  {Basiev}}, \bibinfo {author} {\bibfnamefont {A.~Y.}\ \bibnamefont {Karasik}},
  \bibinfo {author} {\bibfnamefont {V.~V.}\ \bibnamefont {Fedorov}},\ and\
  \bibinfo {author} {\bibfnamefont {K.~W.}\ \bibnamefont {Ver~Steeg}},\
  }\bibfield  {title} {\bibinfo {title} {Optical echo spectroscopy and phase
  relaxation of {Nd}$^{3+}$ ions in {CaF}$_2$ crystals},\ }\href
  {https://doi.org/10.1134/1.558480} {\bibfield  {journal} {\bibinfo  {journal}
  {J. Exp. Theor. Phys.}\ }\textbf {\bibinfo {volume} {86}},\ \bibinfo {pages}
  {156} (\bibinfo {year} {1998})}\BibitemShut {NoStop}%
\bibitem [{\citenamefont {Manson}\ \emph {et~al.}(1995)\citenamefont {Manson},
  \citenamefont {Sellars}, \citenamefont {Fisk},\ and\ \citenamefont
  {Meltzer}}]{Appl0}%
  \BibitemOpen
  \bibfield  {author} {\bibinfo {author} {\bibfnamefont {N.~B.}\ \bibnamefont
  {Manson}}, \bibinfo {author} {\bibfnamefont {M.~J.}\ \bibnamefont {Sellars}},
  \bibinfo {author} {\bibfnamefont {P.~T.}\ \bibnamefont {Fisk}},\ and\
  \bibinfo {author} {\bibfnamefont {R.~S.}\ \bibnamefont {Meltzer}},\
  }\bibfield  {title} {\bibinfo {title} {Hole burning of rare earth ions with
  {{kHz}} resolution},\ }\href {https://doi.org/10.1016/0022-2313(95)00004-A}
  {\bibfield  {journal} {\bibinfo  {journal} {J. Lumin.}\ }\textbf {\bibinfo
  {volume} {64}},\ \bibinfo {pages} {19} (\bibinfo {year} {1995})}\BibitemShut
  {NoStop}%
\bibitem [{\citenamefont {Nilsson}\ \emph {et~al.}(2004)\citenamefont
  {Nilsson}, \citenamefont {Rippe}, \citenamefont {Kr{\"o}ll}, \citenamefont
  {Klieber},\ and\ \citenamefont {Suter}}]{Appl}%
  \BibitemOpen
  \bibfield  {author} {\bibinfo {author} {\bibfnamefont {M.}~\bibnamefont
  {Nilsson}}, \bibinfo {author} {\bibfnamefont {L.}~\bibnamefont {Rippe}},
  \bibinfo {author} {\bibfnamefont {S.}~\bibnamefont {Kr{\"o}ll}}, \bibinfo
  {author} {\bibfnamefont {R.}~\bibnamefont {Klieber}},\ and\ \bibinfo {author}
  {\bibfnamefont {D.}~\bibnamefont {Suter}},\ }\bibfield  {title} {\bibinfo
  {title} {Hole-burning techniques for isolation and study of individual
  hyperfine transitions in inhomogeneously broadened solids demonstrated in
  {Pr}$^{3+}$:{Y}$_2${SiO}$_5$},\ }\href
  {https://doi.org/10.1103/PhysRevB.70.214116} {\bibfield  {journal} {\bibinfo
  {journal} {Phys. Rev. B}\ }\textbf {\bibinfo {volume} {70}},\ \bibinfo
  {pages} {214116} (\bibinfo {year} {2004})}\BibitemShut {NoStop}%
\bibitem [{\citenamefont {Sun}\ \emph {et~al.}(2002)\citenamefont {Sun},
  \citenamefont {Thiel}, \citenamefont {Cone}, \citenamefont {Equall},\ and\
  \citenamefont {Hutcheson}}]{Appl1}%
  \BibitemOpen
  \bibfield  {author} {\bibinfo {author} {\bibfnamefont {Y.}~\bibnamefont
  {Sun}}, \bibinfo {author} {\bibfnamefont {C.~W.}\ \bibnamefont {Thiel}},
  \bibinfo {author} {\bibfnamefont {R.~L.}\ \bibnamefont {Cone}}, \bibinfo
  {author} {\bibfnamefont {R.~W.}\ \bibnamefont {Equall}},\ and\ \bibinfo
  {author} {\bibfnamefont {R.~L.}\ \bibnamefont {Hutcheson}},\ }\bibfield
  {title} {\bibinfo {title} {Recent progress in developing new rare earth
  materials for hole burning and coherent transient applications},\ }\href
  {https://doi.org/10.1016/S0022-2313(02)00281-8} {\bibfield  {journal}
  {\bibinfo  {journal} {J. Lumin.}\ }\textbf {\bibinfo {volume} {98}},\
  \bibinfo {pages} {281} (\bibinfo {year} {2002})}\BibitemShut {NoStop}%
\end{thebibliography}

%

\end{document}